\input{aipcheck}

\documentclass[
    ,final            
  ]
  {aipproc}
\usepackage{epsfig,amssymb,amsmath}
\layoutstyle{6x9}
\pdfoutput=1

\begin{document}

\title{Enhanced standoff sensing resolution using quantum illumination}


\author{Saikat Guha}{
  address={Disruptive Information Processing Technologies group, Raytheon BBN Technologies, \\10 Moulton Street, Cambridge, MA 02138, USA}
}

\author{Jeffrey H. Shapiro}{
  address={Research Laboratory of Electronics, Massachusetts Institute of Technology, \\77 Massachusetts Avenue, Cambridge, MA 02139, USA}
}

\begin{abstract}
Loss and noise quickly destroy quantum entanglement. Nevertheless, recent work has shown that a quadrature-entangled light source can reap a substantial performance advantage over all classical-state sources of the same average transmitter power in scenarios whose loss and noise makes them entanglement breaking~\cite{Sac05, Llo08, Tan08, Guh09, Guh09a, Sha09}, standoff target-detection being an example. In this paper, we make a first step in extending this {\em quantum illumination} paradigm to the optical imaging domain, viz., to obtain better spatial resolution for standoff optical sensing. Our canonical imaging scenario---restricted, for simplicity, to one transverse dimension---is taken to be that of resolving one versus two closely-spaced in-phase specular point targets. We show that an entangled-state transmitter, which uses continuous-wave-pumped spontaneous parametric downconversion (SPDC), achieves an error-probability exponent that exceeds that of all classical-state transmitters of the same average power. Using these error-exponent results, we find the ultimate spatial-resolution limits for coherent-state and SPDC imaging systems that use their respective quantum-optimal receivers, thus quantifying the latter's spatial-resolution advantage over the former. We also propose a structured optical receiver that is ideally capable of harnessing $3$\,dB (of the full $6$\,dB) gain in the error-probability exponent achievable by the SPDC transmitter and its quantum-optimal receiver.
\end{abstract}

\maketitle




Let us consider the task of discriminating between one on-axis point target vs. two off-axis point targets placed at angles $\pm\theta$ from the on-axis direction. The transmitter uses a quasimonochromatic (center frequency $\omega_0$\,rad/s), continuous-wave spontaneous parametric downconversion (SPDC) source, transmitting the signal beam towards the target region. The idler beam is retained, without loss, for subsequent joint measurement with the noisy return from the target region. Each $T$-sec-long transmission comprises $M = WT \gg 1$ signal-idler mode pairs, where $W$ is the source's phase-matching bandwidth. The $M$ signal-idler mode pairs $\{\hat{a}_{S_m},\hat{a}_{I_m}\}$,$1\le m \le M$, are in independent identically-distributed (iid), zero-mean, jointly Gaussian states with the entangled joint state of each mode pair described by the Wigner-distribution covariance matrix
\begin{equation}
{\Lambda}_{SI} = \frac{1}{4}\left[\begin{array}{cccc}
S & C_q & 0 & 0\\
C_q & S & 0 & 0\\
0 & 0 & S & -C_q\\
0 & 0 & -C_q & S
\end{array}\right], 
\end{equation}
where $N_S$ is the average photon number per signal (or idler) mode, $S \equiv 2N_S+1$ and $C_q \equiv 2\sqrt{N_S(N_S+1)}$. The signal beam undergoes an $L$-m vacuum propagation and reflects off the specular point target(s). The return field is incident on a length-$D$ hard-aperture pupil. The roundtrip transmissivity is $\kappa\ll 1$, and $N_B\ge 1$ is the mean noise photon number per mode. The receiver jointly detects and processes the target-return and the retained idler modes in order to make a minimum probability of error (MPE) decision as to which one of the two hypotheses (one versus two targets) is true. Only the signal-bearing spatial modes of the return field, i.e., $\xi_1(x)=\sqrt{1/D}\;{\rm rect}(x/D)$ when there is one target present and $\xi_2(x)=A(k_S, \theta, D)\,\cos(k_S\theta x)\,{\rm rect}(x/D)$ when two targets are present, are relevant to the MPE receiver, where $k_S = 2\pi/\lambda_S$ is the wave number at the signal wavelength $\lambda_S$.  Here, $A(k_S, \theta, D) = \sqrt{{(2/D)}/{(\displaystyle 1+{\rm sinc}(k_S\theta D))}}$ is a normalization constant (with ${\rm sinc}(z) \triangleq {\sin({\pi}z)}/{{\pi}z}$) chosen such that $\bar{n}$, the mean photon number per mode of the field collected by the receiver pupil, is the same under both hypotheses. This normalization ensures that the receiver doesn't base its decision on a signature in the difference in mean photon fluxes under the two hypotheses. Because the target-return spatial modes are \em not\/\rm\ orthogonal, we employ a Gram-Schmidt procedure to construct an orthonormal mode pair $\left\{\phi_1(x),\phi_2(x)\right\}$, with $\phi_1(x) = \xi_1(x)$ and $\phi_2(x)$ proportional to $\cos(k_S\theta{x})-{\rm sinc}(k_S\theta{D}/2)$, which will contain all the relevant target-return information for an MPE decision.  Under hypothesis $H_1$ (one target present), the three modes associated with the return-idler annihilation operators $\{\,\hat{a}_{\phi_{1_m}},\hat{a}_{\phi_{2_m}},\hat{a}_{I_m}\}$ for $1\le m\le M$ are in iid zero-mean jointly Gaussian states with Wigner-distribution covariance matrix
\begin{equation}
\boldsymbol{\Lambda}^{(1)}_{\phi_1\phi_2I} = \frac{1}{4}\left[\begin{array}{cccccc}
A_1 & 0 & C_1 & 0 & 0 & 0 \\
 0 & D_1 & 0 & 0 & 0 & 0\\ 
C_1 & 0 & S & 0 & 0 & 0 \\
0 & 0 & 0 & A_1 & 0 & -C_1 \\
0 & 0 & 0 & 0 & D_1 & 0 \\
 0 & 0 & 0 & -C_1 & 0 & S
\end{array}\right], 
\end{equation}  
where  $D_1 \equiv 2N_B+1$, $A_1 = 2\kappa N_S + D_1$ and $C_1 = \sqrt{\kappa}\,C_q$. Likewise, under hypothesis $H_2$ (two targets present), the modes associated with the annihilation operators $\{\,\hat{a}_{\phi_{1_m}},\hat{a}_{\phi_{2_m}},\hat{a}_{I_m}\}$ for $1\le m\le M$ are in iid zero-mean jointly Gaussian states with Wigner-distribution covariance matrix
\begin{equation}
\boldsymbol{\Lambda}^{(2)}_{\phi_1\phi_2I} = \frac{1}{4}\left[\begin{array}{cccccc}
A_2& B_2 & C_2 & 0 &  0 & 0 \\
B_2 & D_2 & E_2 & 0 & 0 & 0\\ 
C_2 & E_2 & S & 0 & 0 & 0 \\
0 & 0 & 0 & A_2 & B_2 & -C_2 \\
0 & 0 & 0 & B_2 & D_2 & -E_2 \\
 0 & 0 & 0 & -C_2 & -E_2 & S
\end{array}\right],
\end{equation}
$A_2 = 2A^2\kappa N_S + D_1$, $B_2 = 2AB\kappa N_S$, $C_2 = AC_1$, $D_2 = 2B^2\kappa N_S + D_1$ and $E_2 = BC_1$, with
\begin{equation}
A = \sqrt{\frac{2}{1+{\rm sinc}(k_S\theta D)}}\,{\rm sinc}(k_S\theta D/2),  \quad\mbox{and}\quad
B = \sqrt{1 - \frac{2\,{\rm sinc}^2(k_S\theta D/2)}{1 + {\rm sinc}(k_S\theta D)}}.
\end{equation}
Under coherent-state illumination, the pupil-plane return field spatial modes $\left\{\hat{a}_{R_{m}}\right\}$, $1\le m\le M$, are in non-zero-mean thermal states, with $\sqrt{{\rm photons}/{\rm m}}$-unit mean field $\sqrt{\bar n}\;\xi_k(x)$, under hypothesis $H_k$ , for $k =1 ,2$.

\begin{figure}[h]
\centerline{\includegraphics[height=.46\textheight]{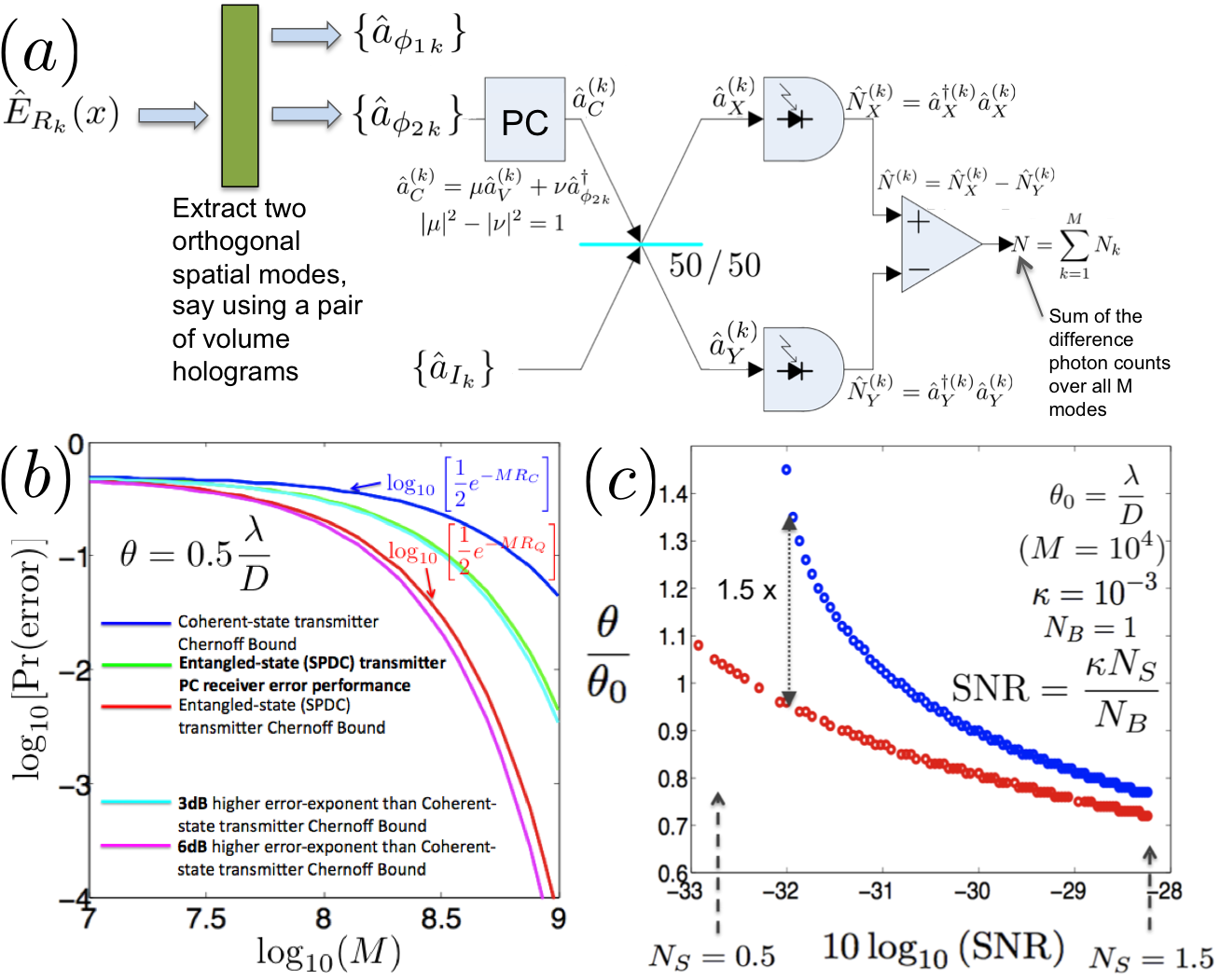}}
\caption{(a) The phase-conjugate (PC) receiver for the one-vs.-two problem, (b) quantum Chernoff bounds for the coherent-state (blue) and SPDC (red) transmitters, and probability of error for the SPDC-transmitter PC-receiver combination (green), versus number of modes $M$, (c) resolution (minimum resolvable angle for a given maximum error probability) vs. received-photon signal-to-noise ratio for the coherent-state (blue) and SPDC (red) transmitters, in each case assuming an optimal receiver.}
\label{fig:results}
\end{figure}

The binary detection problem is the MPE discrimination between hypotheses $H_{1}$ and $H_{2}$ using the optimal measurement on, either, (i) the joint quantum state of the $M$ return modes $({\hat \rho}_{R}^{(k)})^{\otimes M}$ for $k = 1$ or $2$ (for the coherent-state transmitter), or (ii) the joint quantum state of the $3M$ return-idler modes, $({\hat \rho}_{\phi_1\phi_2I}^{(k)})^{\otimes M}$ for $k = 1$ or $2$ (for the quantum-illumination transmitter). The minimum probability of error is given by,
$ P_{e,{\min}}^{(M)} = [1 - \sum_n\gamma_n^{(+)}]/2
$, where $\gamma_n^{(+)}$ are the non-negative eigenvalues of $({\hat \rho}^{(1)})^{\otimes M} - ({\hat \rho}^{(0)})^{\otimes M}$ \cite{Hel76}, where we use ${\hat \rho}^{(k)}$ to denote either ${\hat \rho}_{R}^{(k)}$ (coherent-state transmitter) or ${\hat \rho}_{\phi_1\phi_2I}^{(k)}$ (QI transmitter). The quantum Chernoff bound (QCB), given by $Q_{\rm QCB} \triangleq \min_{0 \le s \le 1}Q_s$ where $Q_s \triangleq {\rm{Tr}}\bigl[({\hat \rho}^{(0)})^s({\hat \rho}^{(1)})^{1-s} \bigr]$, is an exponentially-tight upper bound to $P_{e,{\min}}^{(M)}$ \cite{Aud07}. In particular, we have
\begin{equation}
P_{e,{\min}}^{(M)} \le \frac{1}{2}Q_{\rm QCB}^M \le \frac{1}{2}Q_{0.5}^M,
\label{eq:bounds}
\end{equation}
where the second inequality is the (generally looser) Bhattacharyya bound, and 
$Q_{\rm QCB} = -\lim_{M\to \infty}[\ln(2P_{e,{\rm min}})/M].$
The QCB is often written as $P_{e,{\min}}^{(M)} \le e^{-MR_Q}/2$, where $R_Q \triangleq -{\ln}(Q_{\rm QCB})$ is the error exponent.

The error probability results are shown in Figs.~\ref{fig:results}(b) and~\ref{fig:results}(c). We have chosen $\theta = 0.5\lambda/D$, and have used $N_S=0.01$ mean transmitted photon-number per mode, $N_B=20$ mean noise photons per mode and the roundtrip transmissivity $\kappa=0.01$. Even though we haven't calculated the QCB error exponent analytically, numerical evaluation of the QCB using the procedure outlined in~\cite{Pir08} has revealed---analogous to our results for the target-detection problem~\cite{Tan08}---that the SPDC-transmitter's Chernoff bound is stronger than the coherent-state transmitter's Chernoff bound by 6\, dB (factor of $4$) in the error exponent when $N_S \ll 1$, $\kappa \ll 1$, and $N_B \gg 1$ (see Fig.~\ref{fig:results}(b)). The exponential tightness of the QCB then implies that a  quantum illumination transmitter has a factor-of-four advantage in $-\ln(P_{e_{\rm min}})$ over a coherent-state transmitter when both radiate the same average photon number per mode and both use optimum receivers.   Figure~\ref{fig:results}(c) plots the minimum resolvable target-separation angle as a function of the received signal-to-noise ratio (SNR) ${\kappa}N_S/N_B$, for a $P_e = 0.03$ probability of error threshold. The resolution-versus-SNR plots show the typical divergence behavior:  below a threshold SNR the two targets cannot be resolved.  Quantum illumination's error-exponent advantage enables it to have an appreciably lower threshold SNR than does the coherent-state transmitter.  Thus, for the parameter values chosen ($\kappa = 10^{-3}$ and $N_B=1$), the resolution plots represent almost a $2$-dB lateral shift along the SNR axis. For higher $N_B$ values, quantum illumination's SNR advantage is more pronounced. 

What these results do not tell us yet, is a structured realization of the receiver (in conjunction with the SPDC transmitter) that would be able to harness this $6$\,dB error exponent advantage over the coherent-state transmitter. Figure~\ref{fig:results}(a) shows a structured receiver that can bridge half this gap~\cite{Guh09a}, i.e., obtain an error-probability exponent $3$\,dB higher than the coherent state transmitter's optimal error exponent. The phase-conjugate (PC) receiver conjugates the return modes ${\hat a}_{\phi_{2m}}$ and mixes them with corresponding idler modes ${\hat a}_{I_m}$ on a 50-50 beam splitter. Balanced detection of the resulting ouputs, as shown in Fig.~\ref{fig:results}(a), and summing over all $M$ modes, then provides a sufficient statistic for deciding between the two hypotheses. Quantum illumination with the PC receiver can be shown, analytically, to enjoy a 3\,dB error-exponent advantage over the coherent-state system when $N_S \ll 1$, $N_B \gg 1$, $\kappa \ll 1$, and $\theta \to 0$. Extending {\em quantum illumination} to more complex imaging scenarios, and building a structured design of the quantum-optimal receiver are subjects of ongoing investigation.

\vspace{7pt}
This work was supported by the DARPA Quantum Sensors Program.



\bibliographystyle{aipproc}   





\end{document}